\documentclass{elsarticle}

\usepackage{hyperref,comment,color,float}
\usepackage[figuresright]{rotating}
\usepackage{multirow}
\usepackage{amssymb}
\usepackage[export]{adjustbox}
\usepackage{subfig}
\usepackage{graphicx}
\journal{}

\begin{document}

\begin{frontmatter}

\title{A novel approach to quantify volatility prediction}

\author[1]{Suchetana Sadhukhan}
\ead{s.suchetana@gmail.com}
\author[2]{Shiv Manjaree Gopaliya}
\ead{shivmanjree@gmail.com}
\author[2]{Pushpdant Jain}
\ead{pushpdant@gmail.com}
\address[1]{Department of Physics, School of Advanced Sciences and Languages, VIT Bhopal University, Kothri Kalan, Sehore-466114, India}
\address[2]{School of Mechanical Engineering, VIT Bhopal University, Kothri Kalan, Sehore-466114, India}


\address{}

\begin{abstract}
Volatility prediction in the financial market helps to understand the profit and involved risks in investment. However, due to irregularities, high fluctuations, and noise in the time series, predicting volatility poses a challenging task. 
In the recent Covid-19 pandemic situation, volatility prediction using complex intelligence techniques has attracted enormous attention from researchers worldwide. In this paper, a novel and simple approach based on the robust least squares method in two approaches a) with least absolute residuals (LAR) and b) without LAR, have been applied to the Chicago Board Options Exchange (CBOE) Volatility Index (VIX) for a period of ten years. For a deeper analysis, the volatility time series has been decomposed into long-term trends, and seasonal, and random fluctuations. The data sets have been divided into parts viz. training data set and testing data set. The validation results have been achieved using root mean square error (RMSE) values. It has been found that robust least squares method with LAR approach gives better results for volatility (RMSE = 0.01366) and its components viz. long term trend (RMSE = 0.10087), seasonal (RMSE = 0.010343) and remainder fluctuations (RMSE = 0.014783), respectively. For the first time, generalized prediction equations for volatility and its three components have been presented. Young researchers working in this domain can directly use the presented prediction equations to understand their data sets. 
\end{abstract}

\begin{keyword}
volatility, prediction, kolmogorov zurbenko filter, LAR method, financial market


\end{keyword}

\end{frontmatter}


\section{Introduction}
With the present Covid-$19$ pandemic situation across the globe, accurate prediction and analysis of the stock market have gained much importance  as compared to old times. A lot of research is being focused on accurate stock market prediction. However, volatility prediction in future stock values is a challenging task.  On one side, where the volatility helps in a better understanding of the market risk involved for the investors; on the other side, an accurate prediction model for the same can help in understanding the investment returns. In other words, irrespective of stock market fluctuations, volatility always provides a way to make a profit for investors. Volatility is defined by the random fluctuations in the return prices of an asset in the financial market \cite{pradeepkumar, berry2019}. This aspect was reviewed and a comparative study was carried out on the findings of volatility forecasting published in $93$ papers over a span of two decades\cite{granger2003}. It is well-proved that volatility prediction is difficult to estimate for its dependence on very much random but interdependent variables \cite{damir2021}. Investment risk goes higher for higher volatility. Thus, volatility prediction has gained much attention to assess the market risk and make the proper financial decisions \cite {fangjia2021}. During the present Covid-$19$ pandemic situation, the market has undergone high fluctuations making traditional volatility prediction a very tedious task. To address this issue, the present research has moved towards the application of different hybrid and complex artificial techniques for volatility forecasting. A neural network-based parametric model with idealistic assumptions for volatility forecasting was studied and presented \cite{pranav2021}.  Mettle et al. proposed the method to analyze volatility of an apparently random walk but with correlated errors time series \cite{mettle2022}. This method was further successfully improvised for South Africa (SA), Canada, Swtizerland, and China on their exchange rates to represent the adjudication of SA has the highest variation in exchange rates among all and thus regarded as the most unstable economy. The financial risk management of an organization was predicted and examined by Li et al. \cite{li2022} using a logistic regression model, which gave better results than the conventional method. The effectiveness of the proposed method was proved experimentally to predict the financial data risks. As an advancement in this direction, a framework of deep learning model was adopted for longer period price prediction of a stock for a highly volatile market with unpredictable circumstances by Soleymani et al.\cite{soleymani2022}. The effectiveness of the approach was demonstrated experimentally, too. Authors Yazdani et al. have studied the volatility models for stochastic economic markets and associated behavior and correlation on return prices. The proposed algorithm was applied to many financial time-series and achieved a good fit for stock and cryptocurrency markets \cite{yazdani2022}. Authors Jia et al. \cite{fangjia2021} have concluded that the forecasting of volatility in stocks index can be represented in a better way by using the long short-term memory (LSTM) and deep neural network. Deep learning-based models, as well as machine learning have been applied to stock price prediction and forecast of gold volatility \cite{mehtab2021, rahimikia2020, vidal2020}. Authors Berry et al. in $2019$  attempted to determine if the sentiment on stocks from StockTwits micro-blogs can improve volatility prediction \cite{berry2019}. The obtained results were evaluated experimentally. A new validation alongwith selection technique based on the Chan–Karolyi–Longstaff–Sanders (CKLS) method for two currency exchange rates and comparing their performance was proposed by Sikora et al. \cite{sikora2019}. Net regression algorithms with an improved coefficient of determination which helped with forward-moving time frames rather than previous economic indicators to predict volatility over a period of time were proposed \cite{smith2017}. Different econometric methods such as machine learning are widely adopted to predict the instability in the financial market \cite{kapoor2017, ahoniemi2006}. In a similar attempt, other authors have adopted a new model of regression neural network-based particle swarm optimization method to analyze the fluctuations for the financial market \cite{pradeepkumar}. A robust regression model to accurately forecast stock return volatility \cite{he2021}. It was empirically verified that the obtained results were robust to different settings. \\
The existing literature shows that volatility forecasting for the stock market has been attempted with traditional as well as intelligent approaches. After a thorough review, it has been observed that the financial stock market has different scales of motion for strong seasonal or long-term trends which, in turn, complicates the volatility prediction process. Thus, it motivates the authors to observe these patterns and compute the future volatility prediction separately for the different detrended time series. The analysis is based on the CBOE Volatility Index (VIX). The main objectives of this paper are two-folded: Firstly, to detrend the volatility return time-series data into a long-term trend, seasonal and random fluctuation (remainder) components. It is mainly important for market analysis as volatility forecasting highly depends on  the long-term trend, seasonal, and remainder components and for the policymakers to make good decisions depending on the presence of seasonality if any. Secondly, it provides generalized equations to predict future volatility. In this paper, a simple approach to volatility prediction has been proposed. General prediction equations to represent the variations in  the long-term trend, seasonal, and remainder components parameters along with volatility have been presented. The manuscript is organized in following section; Section 2 represents the data utilized for the investigation and the adopted methodologies for detrending the time series into its different component’s future prediction via curve fitting technique. Further, Section 3 depicts a detailed analysis of results along with the discussion. Finally, in section $4$ the authors summarize the findings of the paper.

\section{Data Description and Methodology}
For the time series prediction analysis, here we consider Chicago Board Options Exchange's (CBOE) Volatility Index (VIX), a volatility measure based on S$\&$P $500$ indices from Yahoo finance from $03$-rd January $2011$ to $06$-th May $2022$ ($2857$ trading days). CBOE VIX, introduced in $1993$, is a measure of market expectation of next $1$-month stock market volatility based on Standard and Poor's $500$ stock index option prices. It acts as a sign of market fear (bear) or greed (bull) and thus, helps to predict the overall market's situation based on the movement of VIX.
\subsection{Logarithmic Return}
For the analysis, we consider logarithmic return time-series of volatility indices' daily closure values with $\Delta t= 1$ day for $k$ stocks using the formula,
\begin{equation}
r(t)=\ln \; P (t+\Delta t)- \ln \; P(t)
\label{eqn1}
\end{equation}
Here $r(t)$ and $P(t)$ denote the logarithmic return price and return price, respectively, at time $t$ with $t = 1, 2,...,T$ where $T$ is the number of total trading days. 
\subsection{Kolmogorov-Zurbenko Filter}
It is extensively shown in the literature that real time-series has a long-term trend, short-term seasonal variation, and random fluctuation as its component; financial market \cite{wold1938study, milocs2020multifractal}, air quality data\cite{wise2005, kang2013, li2017, agudelo2014}. There are many methods available for detrending the data viz. PEST Algorithm \cite{brockwell1991stationary}, monthly anomaly technique \cite{wilks2011statistical}, the wave-let transform \cite{lau1995climate, meinl2012, SUN2012, sun2015, berger2016}. To decompose our time series, we have used the Kolmogorov-Zurbenko (KZ) filter proposed by Kolmogorov and Zurbenko in 1986 \cite{zurbenko1986}. The advantage of this method over the others is that it is a simpler method and takes care of the missing data in the time series \cite{eskridge1997separating}. This method separates out the different components to analyze the underlying hidden trends of the time series in a deeper sense \cite{sen2016, eskridge1997separating, yang2010, zurbenko2018}.

An original time-series $A(t)$ can be shown by:
\begin{equation}
A(t)=e(t)+S(t)+W(t)
\end{equation}
where $e(t)$, $S(t)$ and $W(t)$ are the long-term
trend, short-term seasonal, and fluctuation components, respectively. This KZ filter behaves as a low-pass filter during successive iterations of a moving average of window length $m$ and $p$ iterations. It can be depicted as below:
\begin{equation}
Y_i=\frac{1}{m} \sum_{j=-k}^k A_{i+j}
\end{equation}
where, the window length is considered as $m=2k+1$  with $k$ as the number of data points included on both sides of $i$. During each iteration, we get one output, which will be considered as the input for the next iteration. Different window lengths $(m)$ and iteration number $(p)$ find out different scales of motion. 
A $KZ_{(365,3)}$ filter helps in extracting trends of different time-scales or periods from the data. $e(t), s(t), w(t)$ can be calculated for different value of $m$ and $p$  as follows:
\begin{eqnarray}
e(t)=KZ_{(365,3)}\\
S(t)=KZ_{(15,5)}-KZ_{(365,3)}\\
W(t)=A(t)-KZ_{(15,5)}
\label{eqn2}
\end{eqnarray}
\subsection{Curve fitting}
Regression analysis is very important in statistical mathematics analysis. MATLAB-based Curve Fitting method is one of the simple and efficient ways of performing regression for linear as well as non-linear data sets. Literature search highlights the use of this method for prediction purposes in various domains of real-life problems. Two different software viz. MATLAB and LabVIEW for curve fitting purposes were studied and compared which helped to select the appropriate configuration of the software while dealing with the problem associated with the real life \cite{wen2012comparison}. An accurate curve fitting approach for short-term hourly load forecast was proposed \cite{farahat2012}. An artificial neural network (ANN) based curve fitting technique for the prediction of Covid-$19$ cases across the world was also presented \cite{tamang2020forecasting}. The results showed that ANN can predict the possible number of future cases of Covid-$19$ outbreak in any country very efficiently. The authors of this paper observed that the regression analysis using the Curve Fitting method is a popular and widely used analysis tool for many statistical problems, however, no or limited use can be seen in the market fluctuation analysis.  \\

In this paper, regression analysis has been performed using the MATLAB-based Curve Fitting method, a simple yet effective approach to derive general prediction equations for volatility and its three components viz. long-term trend, seasonal, and remainder fluctuation. The Curve Fitting method helps get a curve plot or surface plot for linear and non-linear models. Surface plots help in a better and clear understanding of large data sets such as in this case. The pre-processing and post-processing of data sets, comparative analysis of developed models, and removal of outliers can be done using this method which in turn helps in a better understanding of the data sets under study. MATLAB also provides a platform to exclude outliers and improve the goodness of fit. This feature has been suitably used wherever required. The non-parametric modeling technique such as splines, interpolation, and smoothing is also possible with the MATLAB-based Curve Fitting method. The coefficients are estimated through the least-squares method which minimizes the summed square of residuals. Residuals ($r_i$), or the errors associated with the data for the $i$-th data point, are defined as the difference between the observed $y_i$ and fitted response value $\overline{y_i}$.
Residual = Data – Fit\\
\begin{eqnarray}
r_i=y_i-\overline{y_i}
\label{eqn3}
\end{eqnarray}
The summed square of residuals is given by
\begin{eqnarray}
S=\sum_{i=1}^n {r_i}^2=\sum_{i=1}^n (y_i-\overline{y_i})^2
\label{eqn4}
\end{eqnarray}
Here $n$ and $S$ represents the number of data points in the fit and sum of squared error estimate respectively. It was specified that response errors represent a normal distribution consisting of rare extreme values i.e., outlier. The main disadvantage of the least-squares fitting is its sensitivity to these values as the squaring of the residuals magnifies the effects of extreme data points. In order to reduce the effect, the outliers are removed using residual plots which have helped to improve the results further. A robust least-squares regression has also been adopted during the present research work. Two approaches of robust regression method for fitting can be applied \cite{gopaliya2020prediction}:
\begin{itemize}
    \item Least absolute residuals (LAR): \\
    It computes a curve by minimizing the absolute difference of the residuals. As it does not consider the squared differences in the calculation, it makes advantage of extreme values having a lesser influence on the fit.
    \item Bi-square weights: \\
    This method uses a weighted sum of squares and minimizes the same. The weight for each data point is a function of the distance of the point from the fitted line. More or full weight will be given to the points near the line. As points move farther from the line, the weight gets reduced. 
    \end{itemize}
Volatility prediction is a very important task due to random fluctuations in the market. In this paper, the robust least squares method has been used for volatility prediction. This paper has focused on volatility and its components viz. long-term trend, seasonal, and remainder fluctuation over a decade from $2011$ to $2022$. During the present study, it was found that the comparable goodness of fit could only be obtained for volatility and its variables through the LAR method. General prediction equations based on stock market parameters viz. volatility and volatility as a function of the long-term trend, seasonal, and remainder fluctuation components have been derived using above discussed Curve-Fitting methods. 

\section{Results and discussion}
\begin{figure*}
	\centering
	\includegraphics[width=0.5\textwidth,height=0.5\textwidth]{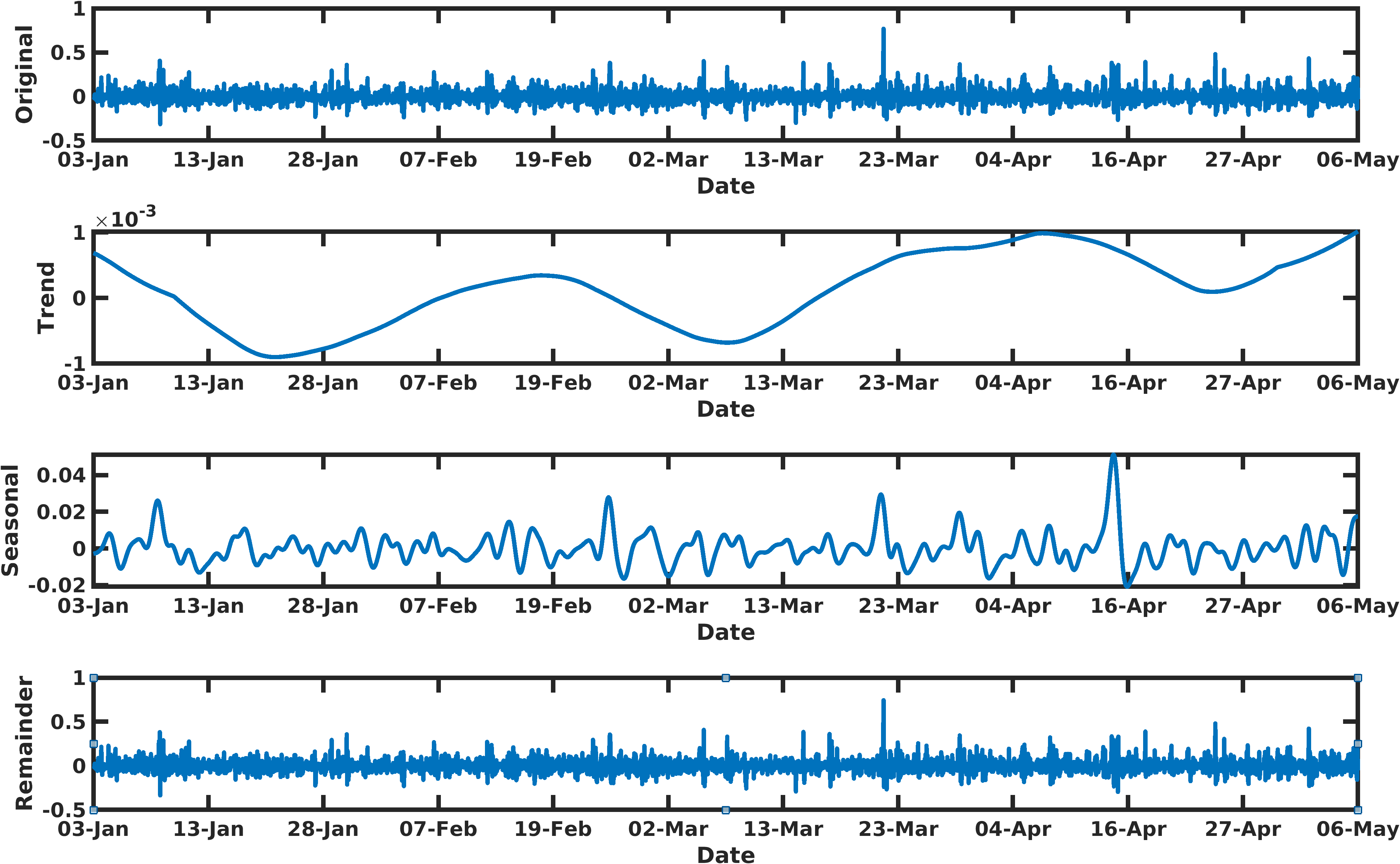}
\caption{Figure represents the original times series considered for the analysis and its decomposition to the long-term trend, seasonal trend, and the remaining fluctuation component as a remainder, respectively, from top to bottom}
\label{fig1}
\end{figure*}
For the prediction, we have considered the time series of CBOE Volatility Index (VIX) daily closure return prices from $03$-rd January $2011$ to $06$-th May $2022$ ($2857$ trading days). We have started by computing the logarithmic return [see Eqn \ref{eqn1}] which is a very common practice for researchers to normalize the data set. For further decomposition to different scales of motion, we applied KZ filter to the logarithmic return volatility time series. Figure \ref{fig1} shows the logarithmic return volatility time series (top) and its decomposition to the long-term trend, seasonal trend, and remainder components from top to bottom, respectively. For prediction purposes, we have considered the original time series and its three components as separate time series. Over the span of $2011$ to $2022$, a total of $2857$ data points have been considered, where the first $2000$ points have been used as training data set while the next $857$ points have been used to test the effectiveness of the used approach. In the present work, the robust least-squares method in two ways viz. a) without LAR and b) with LAR has been used for regression analysis of volatility prediction. The validation results have been achieved by observing the root mean square error (RMSE) values. RMSE gives the average distance between the predicted values from the regression model and the actual values of the data sets. Lower RMSE values, nearing 0, have been accepted as it implies a better fit of the data set.\\


In the robust least squares Curve Fitting method, the training data set points have been used to get three-dimensional surface plots and residual plots for volatility and its three components. A surface plot helps in a better understanding of large data sets. A residual plot graphically represents the difference between predicted values and actual values of the data set. It also helps in omitting the outliers which in turn helps in getting better prediction results. For the first time as a part of the present research work, general prediction equations have been obtained for volatility and its three components. Young researchers can directly use our general prediction equations to perform market volatility fluctuation analysis for their future work.\\
Generalized equation is of the form:$f(x,y)=p_{mn}*x^m*y^n$ with $m$ and $n$ as the coefficients. Detailed values are given in Table \ref{tab1} and \ref{tab2} for without LAR and with LAR, respectively.
\begin{table}[]
\resizebox{\textwidth}{!}{%
\begin{tabular}{|l|l|l|l|l|}
\hline
m 			(rightwards)/ n(downwards) &  & 0 & 1 & 2 \\ \hline
\multirow{4}{*}{0} & V & 0.01233 			(-0.01694, 0.04161) & 0.0281 			(0.02165, 0.03456) & -0.0004441 			(-0.0006191, -0.000269) \\ \cline{2-5} 
  & T & 8.672 			(4.76, 12.58); & -0.6301 			(-1.988, 0.7277) &  \\  \cline{2-5}
 & S & 4.047 			(1.267, 6.827); & 0.1276 			(-0.6494, 0.9046) & -0.03172 			(-0.05457, -0.00886) \\ \cline{2-5} 
  & R & -0.03512 			(-0.1151, 0.04484) & 0.08812 			(0.06091, 0.1153) & 0.001183 			(2.387e-05, 0.002341) \\\hline
\multirow{4}{*}{1} & V & -0.03204 			(-0.03979, -0.0243) & 0.0005061 			(0.0002903, 0.000722) &  \\ \cline{2-5} 
 & T & -1.361 			(-3.217, 0.496); & 0.1253 			(-0.1432, 0.3939) &  \\ \cline{2-5}
 & S & -0.5251 			(-1.464, 0.4142); & 0.1438 			(0.07457, 0.213) &  \\ \cline{2-5} 
  & R & -0.09067 			(-0.124, -0.05736) & -0.009049 			(-0.01266, -0.005441); & -4.646e-05 			(-9.759e-05, 4.665e-06); \\ \hline
\multirow{4}{*}{2} & V &  &  &  \\ \cline{2-5} 
 & T & 0.04688 			(-0.2704, 0.3642) & -0.007231 			(-0.02617, 0.01171) &  \\ \cline{2-5} 
 & S & -0.1123 			(-0.1674, -0.0571) &  &  \\ \cline{2-5} 
  & R & 0.008189 			(0.004919, 0.01146) & 0.0002234 			(9.637e-05, 0.0003505) &  \\ \hline
\multirow{4}{*}{3} & V &  &  &  \\ \cline{2-5}
 & T & -4.575e-05 			(-0.02162, 0.02153); & 0.0001691 			(-0.000396, 0.0007342) &  \\ \cline{2-5} 
  & S &  &  &  \\ \cline{2-5} 
 & R & -0.0001844 			(-0.000274, -9.476e-05) &  &  \\ 
 \hline
\multirow{4}{*}{4} & V &  &  &  \\ \cline{2-5} 
 & T & -1.937e-05 			(-0.0006531, 0.0006144); & -1.427e-06 			(-7.473e-06, 4.619e-06 &  \\ \cline{2-5}
  & S &  &  &  \\ \cline{2-5} 
 & R &  &  &  \\  
 \hline

\multirow{4}{*}{5} & V &  &  &  \\ \cline{2-5} 
 & T & 2.342e-07 			(-6.487e-06, 6.955e-06) &  &  \\  \cline{2-5} 
  & S &  &  &  \\ \cline{2-5} 
 & R &  &  &  \\ \hline
\end{tabular}%
}
\caption{Table presents the various coefficient values (with $95$\% confidence bounds) for the generalized equation $f(x,y)=p_{mn}*x^m*y^n$ with $m$ and $n$ as the coefficients for without LAR method. Abbreviations V, T, S, and R represent the Volatility, Trend, Seasonal, and, Remainder respectively. The range given within the bracket beside the main coefficient value defines the allowed range value of the fitting. }
\label{tab1}
\end{table}

\begin{table}[]
\resizebox{\textwidth}{!}{%
\begin{tabular}{|l|l|l|l|l|}
\hline
m 			(rightwards)/ n(downwards) &  & 0 & 1 & 2 \\ \hline
\multirow{4}{*}{0} & V & -0.0005495 			(-0.001239, 0.00014) & 0.0708 			(0.06661, 0.07499) & -0.008997 			(-0.01013, -0.007866) \\ \cline{2-5} 
 & T & -0.0001798 			(-0.0001879, -0.0001717); & 0.001243 			(0.001199, 0.001287) &  \\ \cline{2-5}
  & S & -0.0002757 			(-0.0003326, -0.0002187) & 0.006036 			(0.005691, 0.00638) & -0.001052 			(-0.001182, -0.0009216) \\ \cline{2-5} 
 & R & 0.0009461 			(-1.202e-05, 0.001904) & 0.06214 			(0.05675, 0.06753) & 0.05465 			(0.05134, 0.05796) \\  

 \hline
\multirow{4}{*}{1} & V & -0.06842 			(-0.07254, -0.0643) & 0.008747 			(0.007568, 0.009926) &  \\ \cline{2-5} 
 & T & -0.001264 			(-0.00131, -0.001218) & 0.0001088 			(5.642e-05, 0.0001613) &  \\  \cline{2-5}
 & S & -0.006055 			(-0.006394, -0.005716) & 0.004716 			(0.004383, 0.00505); &  \\ \cline{2-5}
 & R & -0.06164 			(-0.06695, -0.05633) & -0.1476 			(-0.1566, -0.1387) & -0.01962 			(-0.02162, -0.01761) \\ \hline
\multirow{4}{*}{2} & V &  &  &  \\ \cline{2-5} 
 & T & 1.45e-05 			(-3.725e-05, 6.624e-05) & 0.0007196 			(-0.0007818, -0.0006573) & \\ \cline{2-5} 
  & S & -0.003453 			(-0.003677, -0.003229) &  &  \\ \cline{2-5} 
 & R & 0.0874 			(0.08061, 0.09419) & 0.0426 			(0.03839, 0.04681) &  \\
\hline
\multirow{4}{*}{3} & V &  &  &  \\ \cline{2-5} 
 & T & 0.0006142 			(0.0005536, 0.0006747) & 0.0002639 			(0.0002295, 0.0002983) &  \\ \cline{2-5}
  & S &  &  &  \\ \cline{2-5} 
 & R & -0.02191 			(-0.02441, -0.0194) &  &  \\  
\hline
\multirow{4}{*}{4} & V &  &  &  \\ \cline{2-5}
 & T & -0.0002303 			(-0.0002634, -0.0001971) & -2.714e-05 			(-3.233e-05, -2.194e-05) &  \\  \cline{2-5} 
  & S &  &  &  \\ \cline{2-5} 
 & R &  &  &  \\
 \hline
\multirow{4}{*}{5} & V &  &  &  \\ \cline{2-5} 
 & T & 2.37e-05 			(1.876e-05, 2.864e-05) &  &  \\ \cline{2-5}
 & S &  &  &  \\ \cline{2-5} 
 & R &  &  &  \\  
\hline
\end{tabular}%
}
\caption{Table presents the various coefficient values (with $95$\% confidence bounds) for the generalized equation $f(x,y)=p_{mn}*x^m*y^n$ with $m$ and $n$ as the coefficients for with LAR method. Abbreviations V, R, S, and T represent the volatility, Remainder, Seasonal, and Trend, respectively. The range given within the bracket beside the main coefficient value defines the allowed range value of the fitting. }
\label{tab2}
\end{table}

\begin{figure}[!htp]
  \subfloat[]{\label{fig1:a}\includegraphics[width=0.5\textwidth]{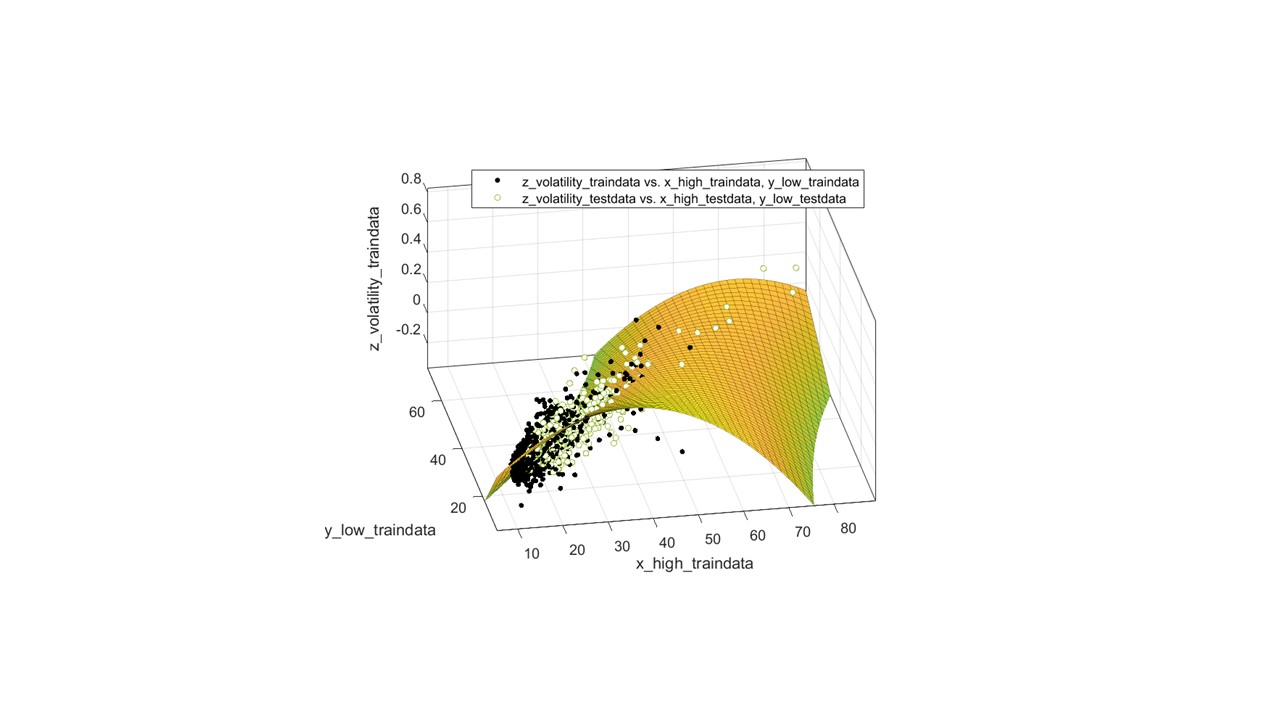}}
  \subfloat[]{\label{fig1:b}\includegraphics[width=0.5\textwidth]{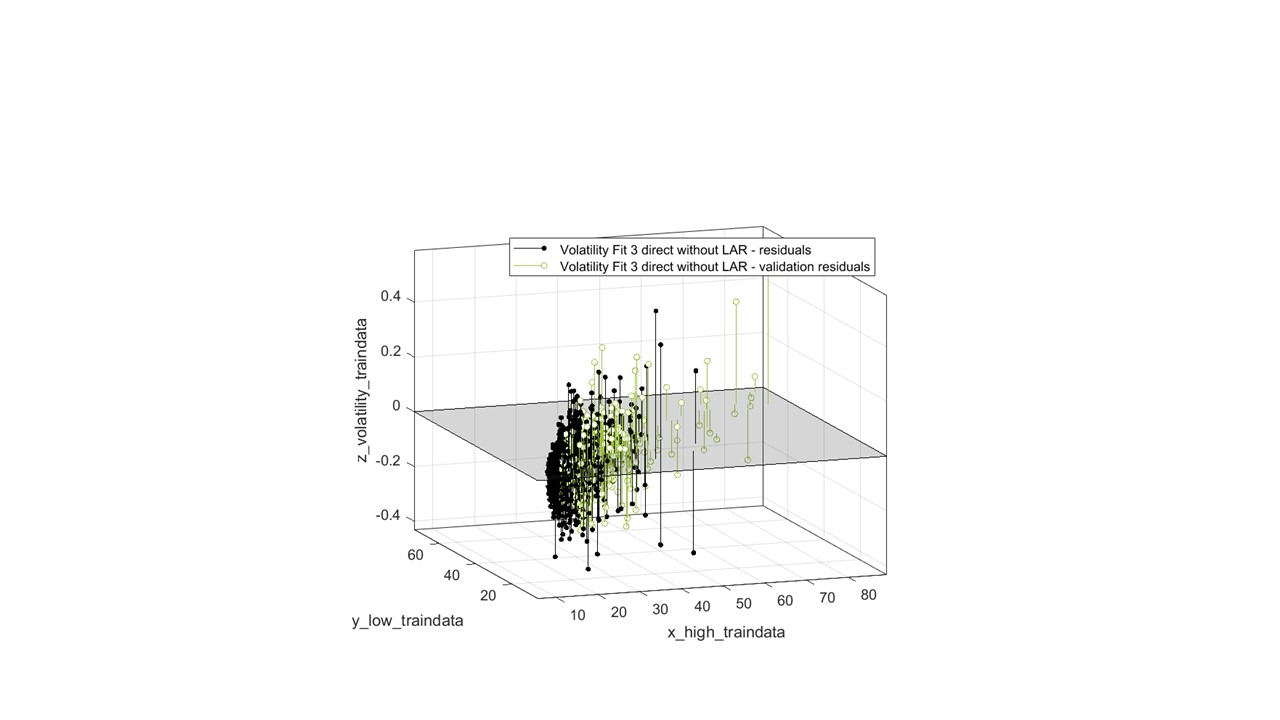}}\\
  \subfloat[]{\label{fig1:a}\includegraphics[width=0.5\linewidth]{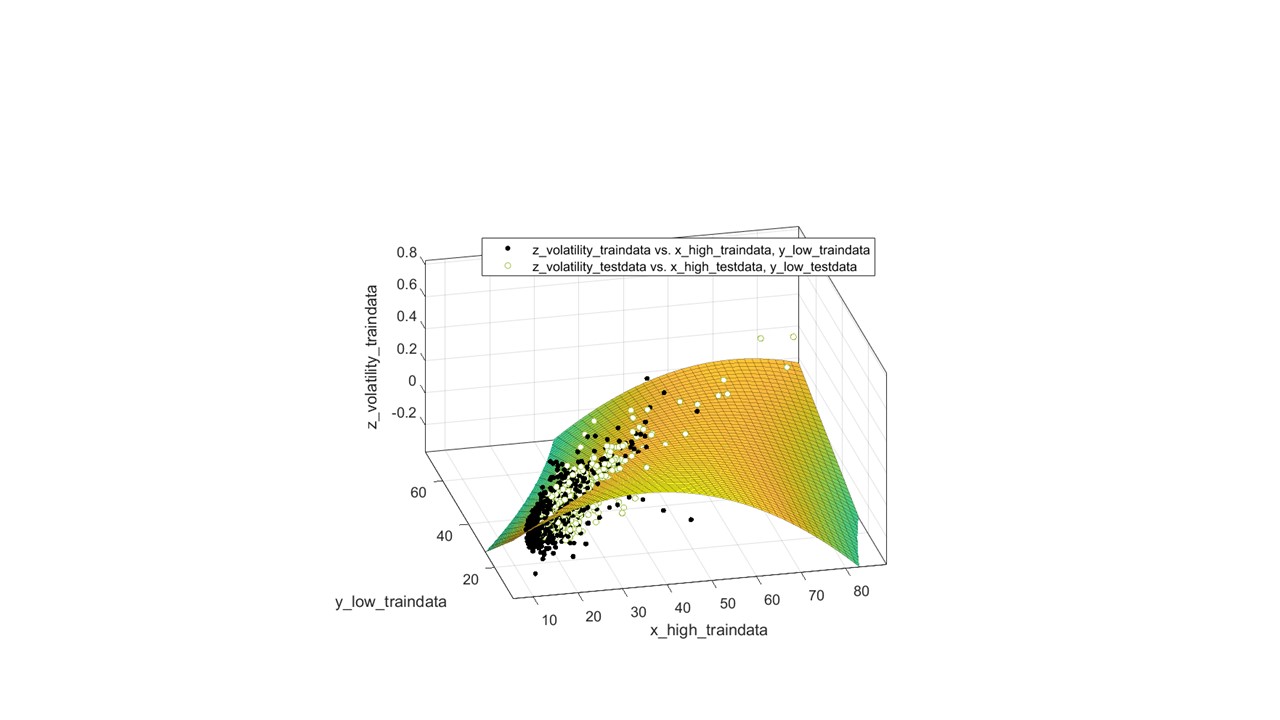}}
  \subfloat[]{\label{fig1:b}\includegraphics[width=0.5\linewidth]{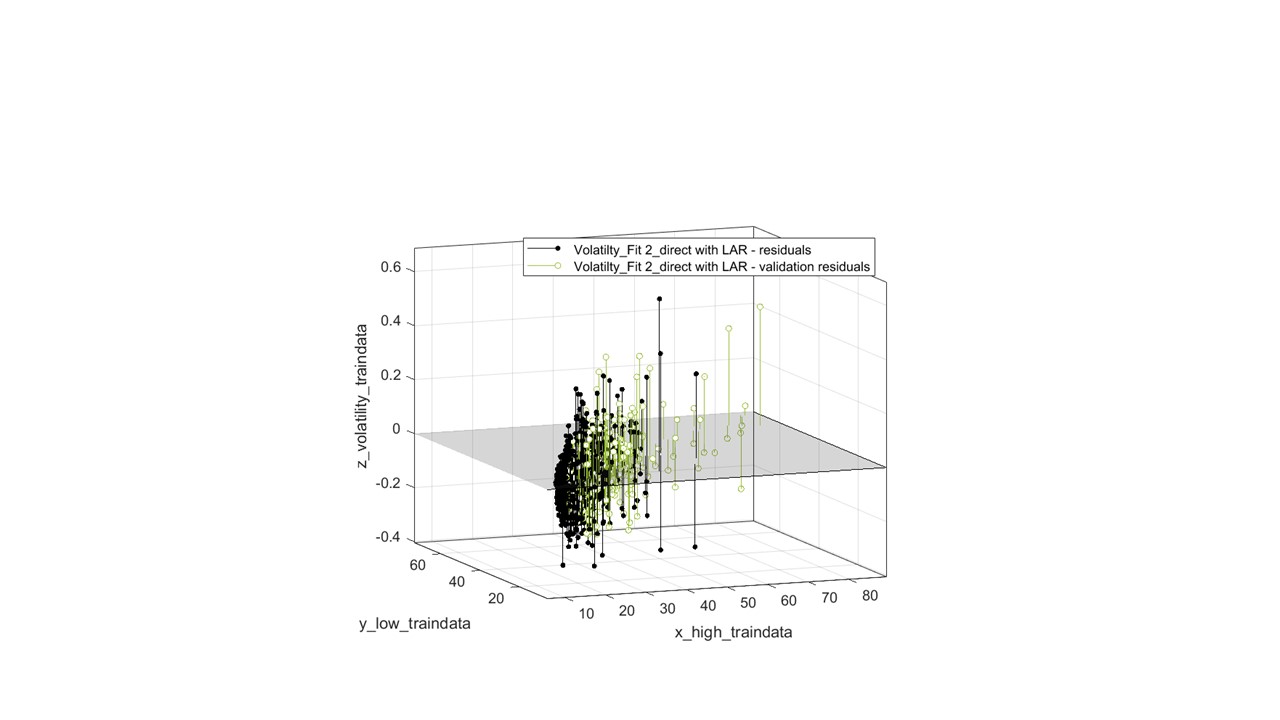}}
  	\caption{Volatility prediction using robust least squares method (a and b): without LAR and (c and d): with LAR; left: Surface plots, right: Residual plots}
\label{fig2}
\end{figure}

\begin{figure}[!htp]
  \subfloat[]{\label{fig1:a}\includegraphics[width=0.5\textwidth]{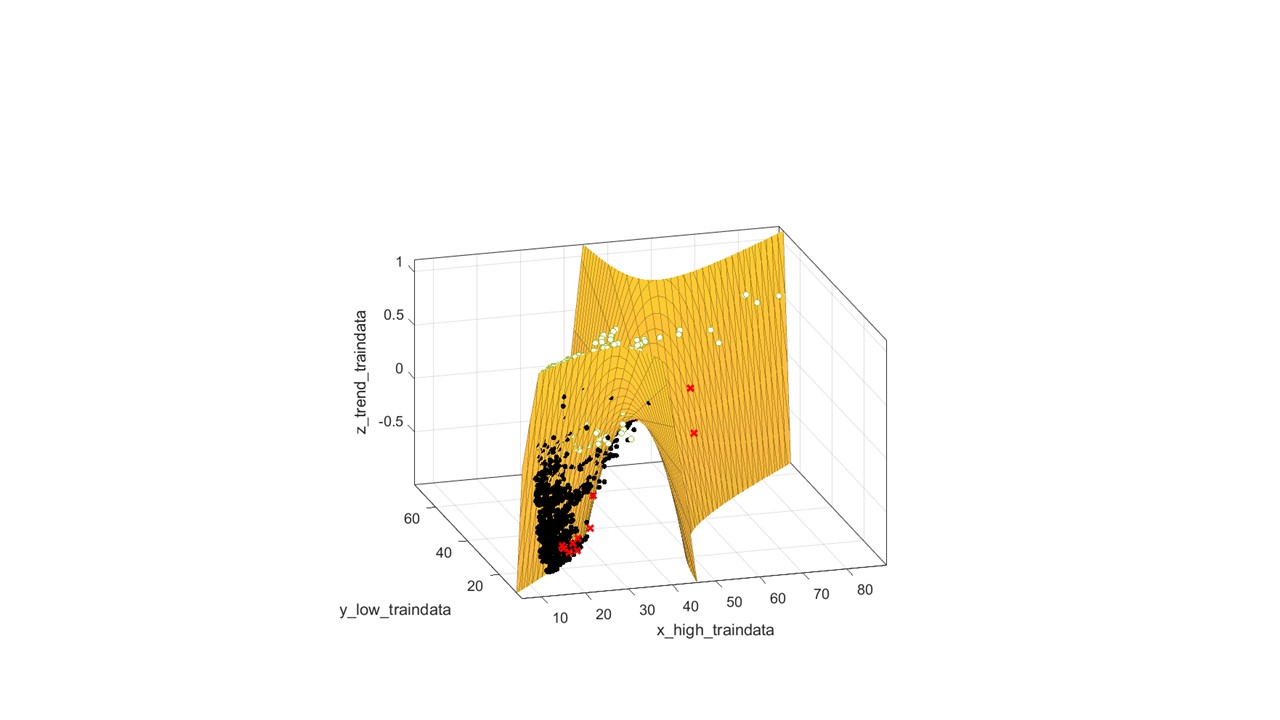}}
  \subfloat[]{\label{fig1:b}\includegraphics[width=0.5\textwidth]{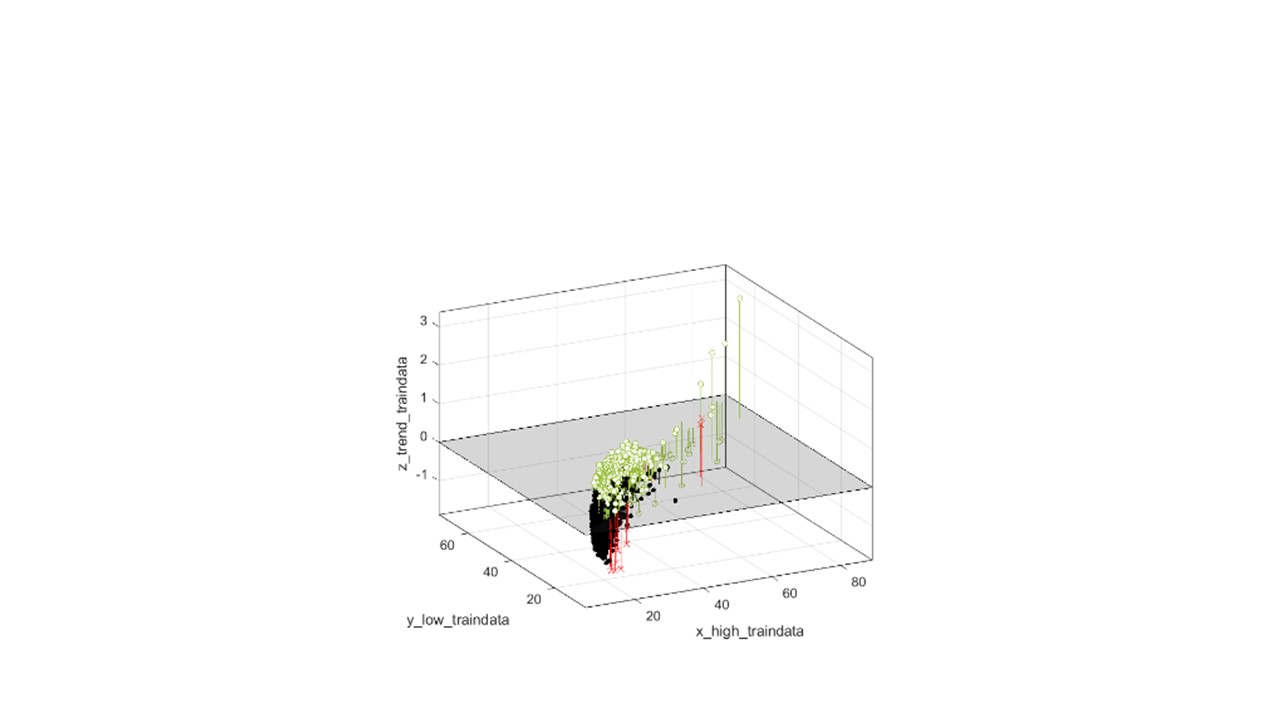}}\\
  \subfloat[]{\label{fig1:a}\includegraphics[width=0.5\linewidth]{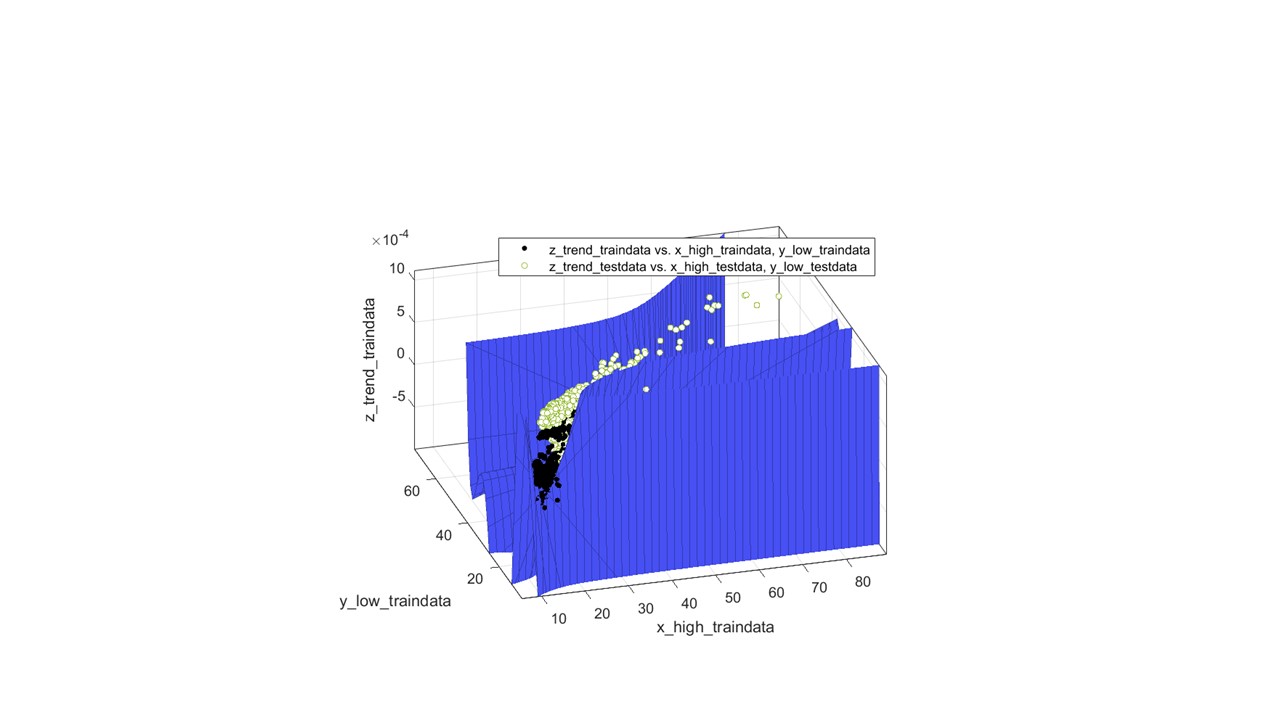}}
  \subfloat[]{\label{fig1:b}\includegraphics[width=0.5\linewidth]{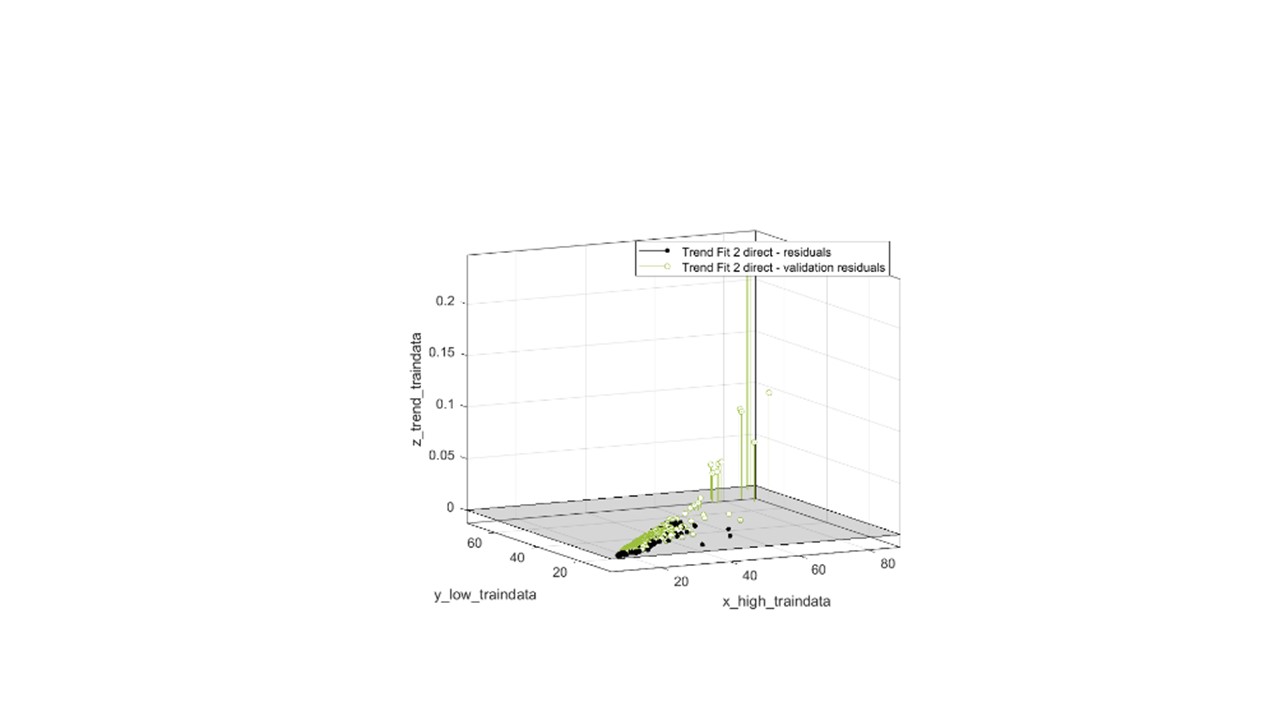}}
\caption{Long-term trend prediction using robust least squares method (a and b): without LAR and (c and d): with LAR; left: Surface plots, right: Residual plots}
\label{fig3}
\end{figure}
\begin{figure}[!htp]
  \subfloat[]{\label{fig1:a}\includegraphics[width=0.5\textwidth]{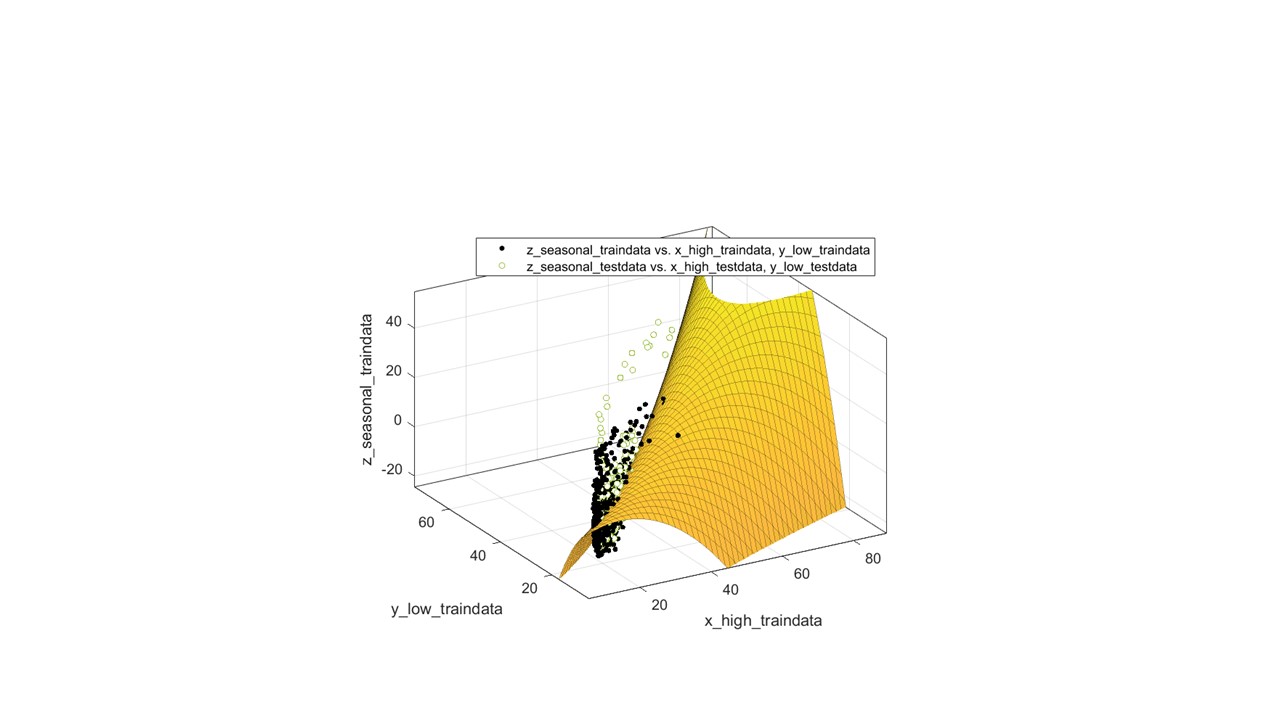}}
  \subfloat[]{\label{fig1:b}\includegraphics[width=0.5\textwidth]{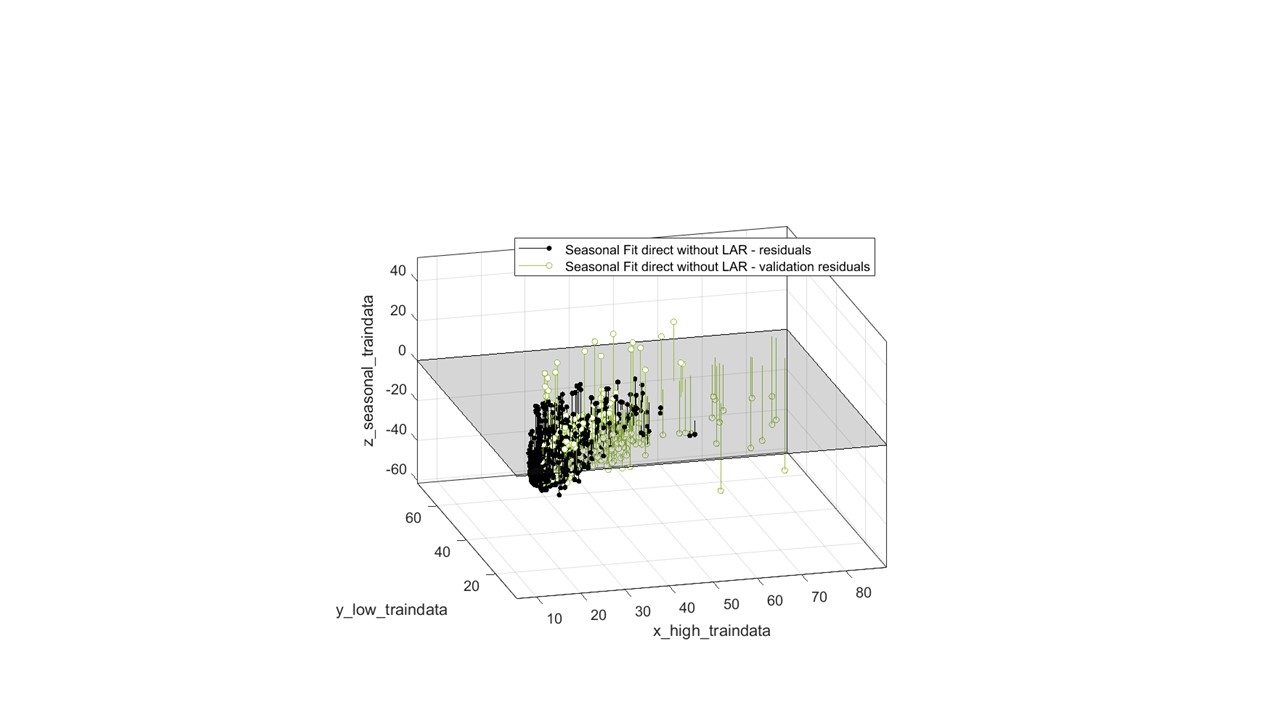}}\\
  \subfloat[]{\label{fig1:a}\includegraphics[width=0.5\linewidth]{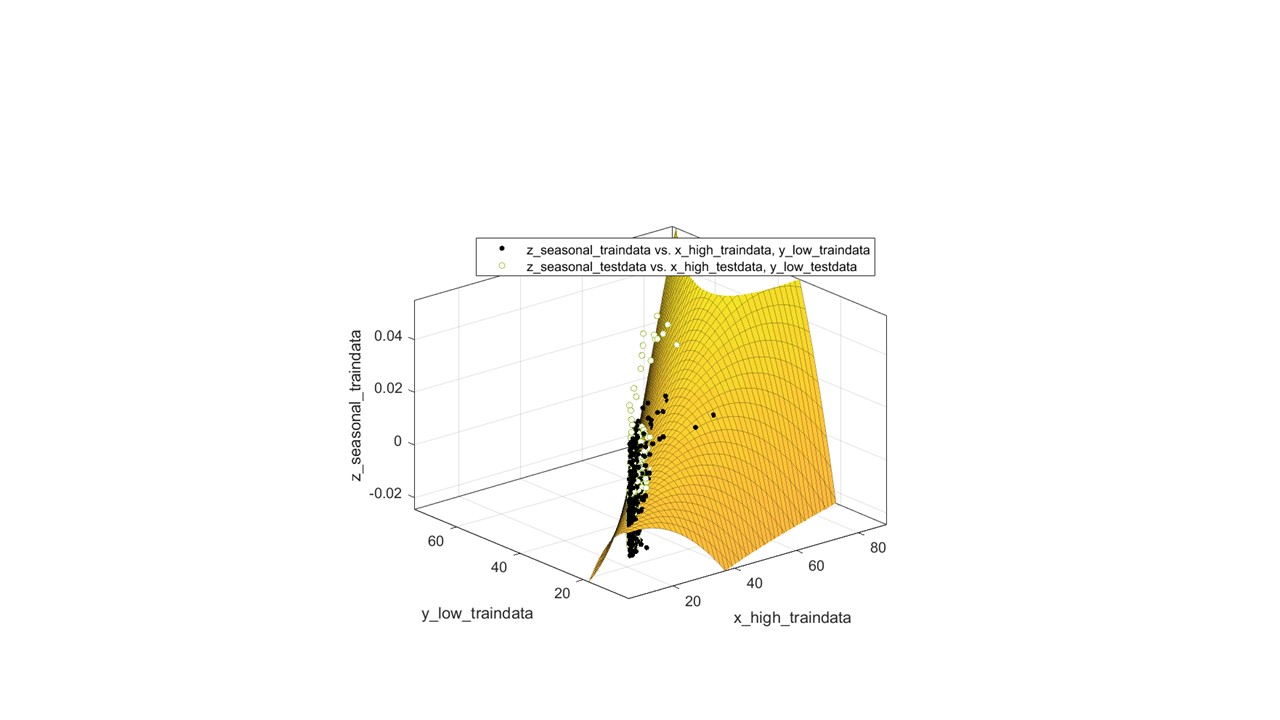}}
  \subfloat[]{\label{fig1:b}\includegraphics[width=0.5\linewidth]{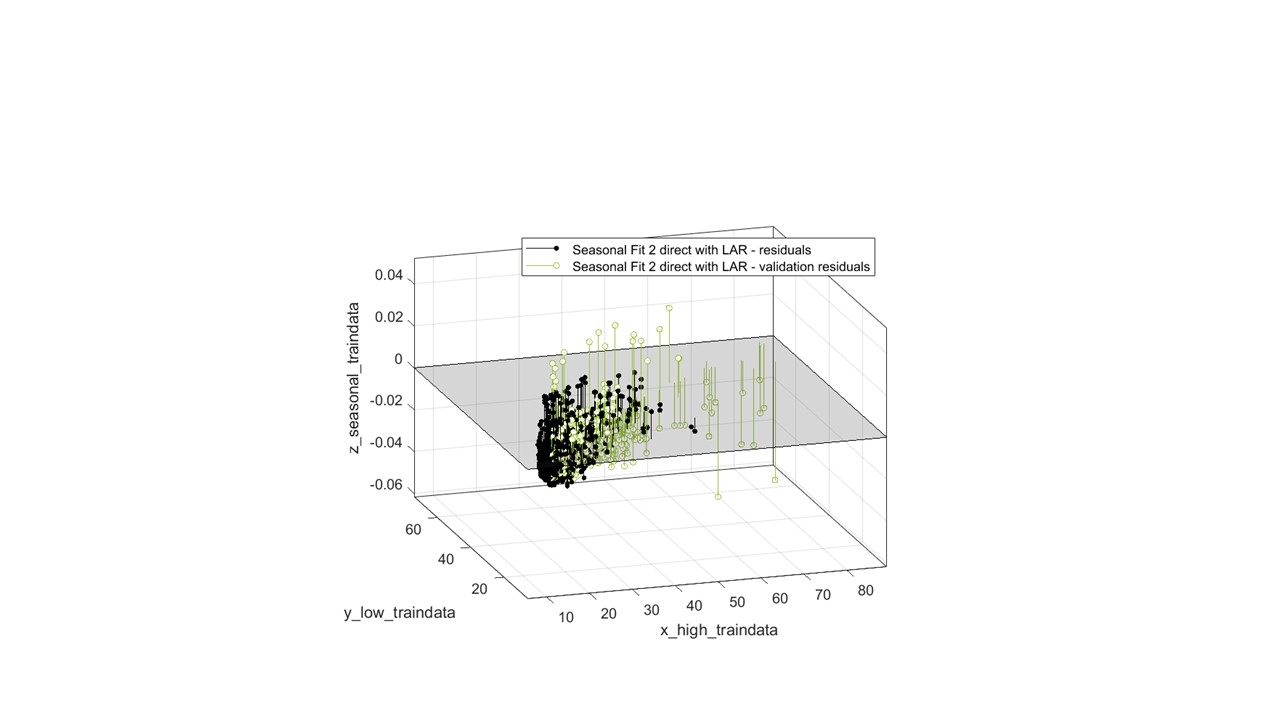}}
	\caption{Seasonal prediction using robust least squares method (a and b): without LAR and (c and d): with LAR; left: Surface plots, right: Residual plots}
\label{fig4}
\end{figure}
\begin{figure}[!htp]
  \subfloat[]{\label{fig1:a}\includegraphics[width=0.5\textwidth]{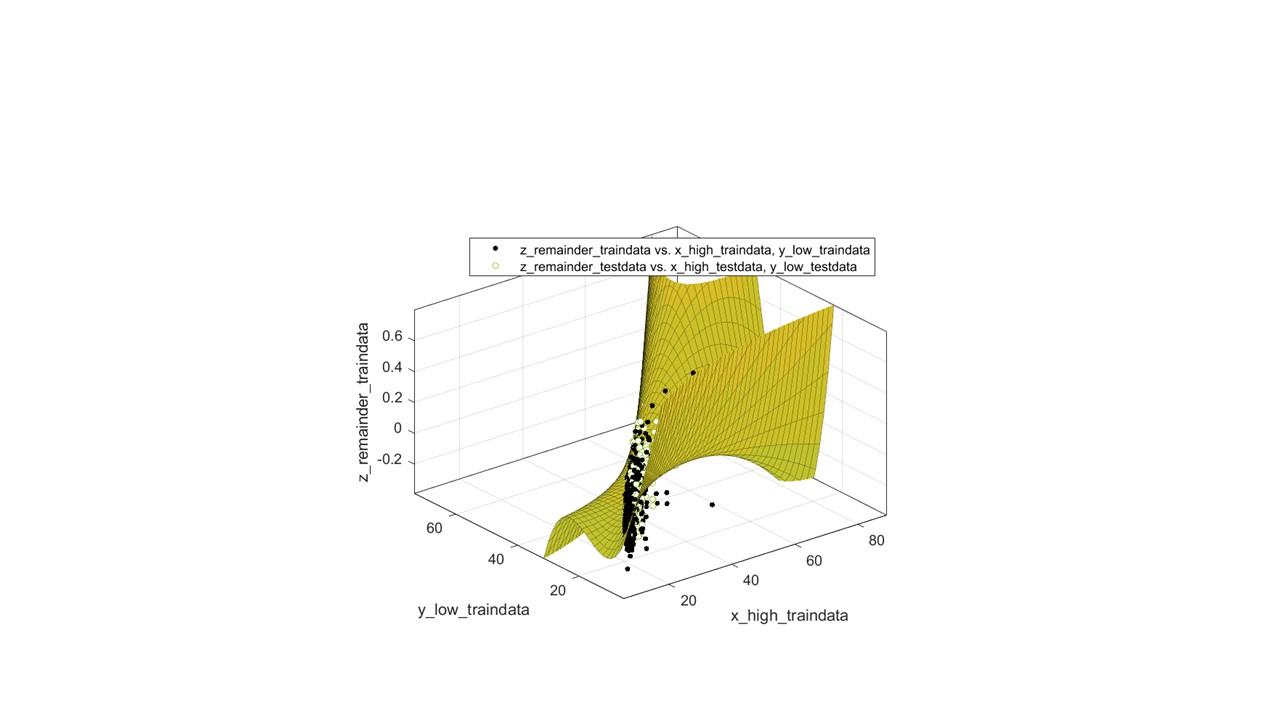}}
  \subfloat[]{\label{fig1:b}\includegraphics[width=0.5\textwidth]{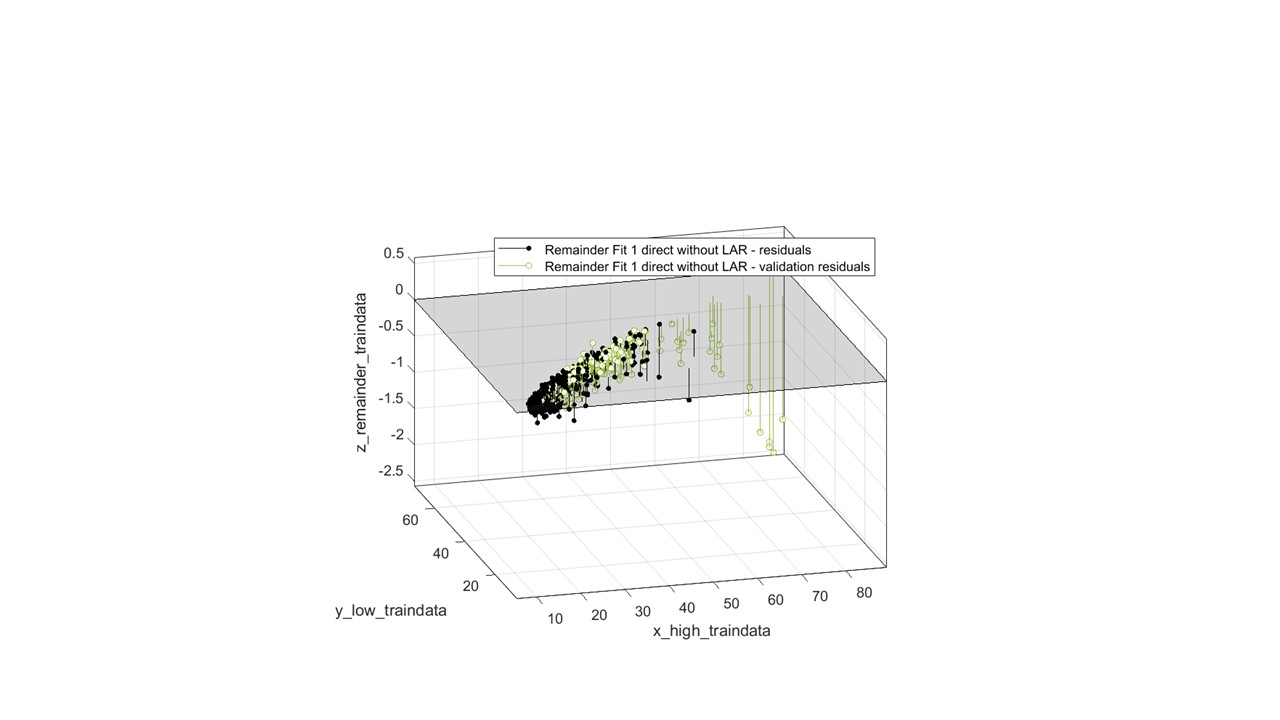}}\\
  \subfloat[]{\label{fig1:a}\includegraphics[width=0.5\linewidth]{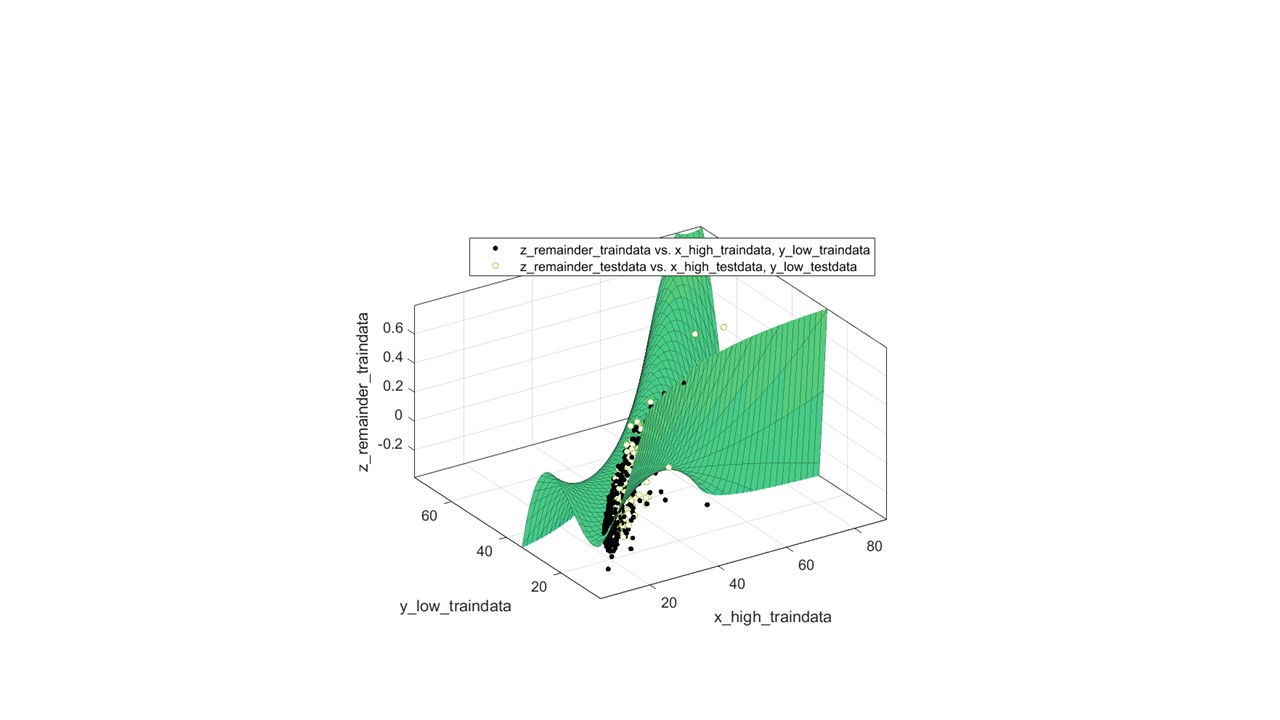}}
  \subfloat[]{\label{fig1:b}\includegraphics[width=0.5\linewidth]{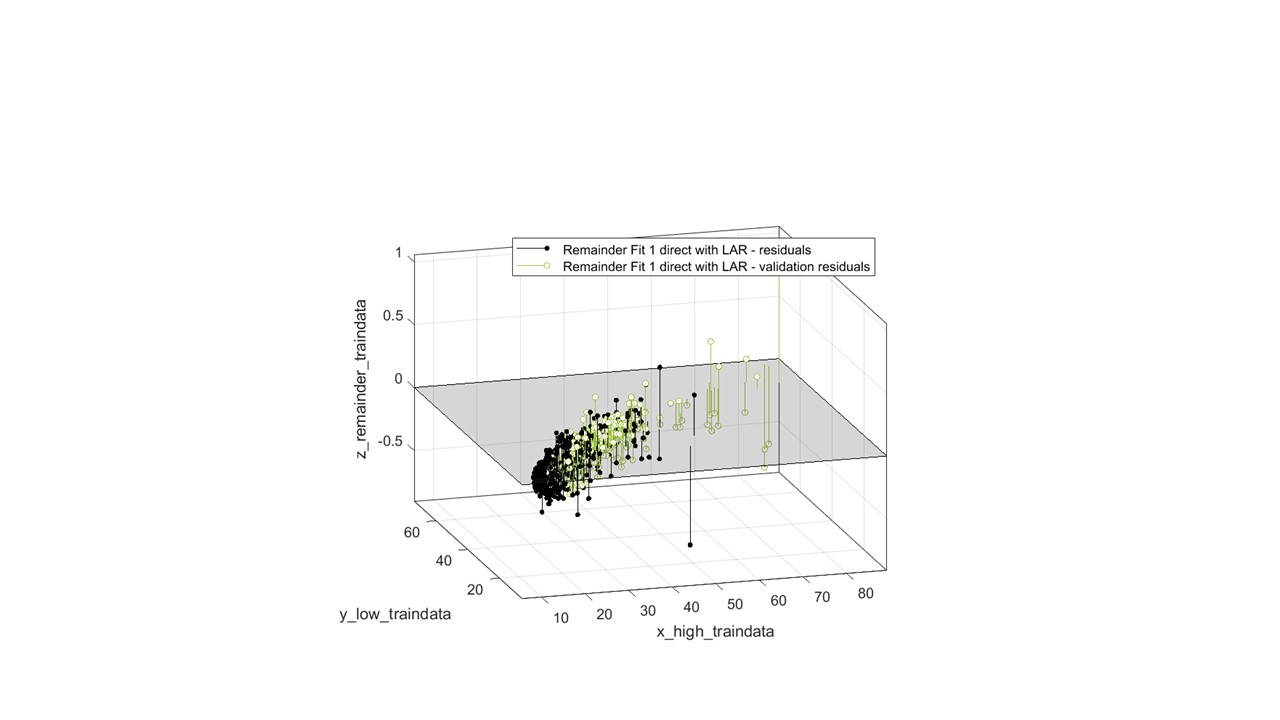}}
  	\caption{Remainder fluctuations prediction using robust least squares method (a and b): without LAR and (c and d): with LAR; left: Surface plots, right: Residual plots}
\label{fig5}
\end{figure}

Figure \ref{fig2} (a and b) shows the surface plot and residual plot for volatility prediction without the LAR method while Figure \ref{fig2} (c and d) shows the surface plot and residual plot for volatility with the LAR method. It can be deduced that volatility prediction has better results with LAR (RMSE = $0.01366$) than without LAR (RMSE = $0.07559$) methods for training data sets. However, tested data sets show  similar prediction results for volatility without LAR (RMSE = $0.0811392$) and with LAR (RMSE = $0.0818176$).\\

Figure  \ref{fig3} (a and b) shows the surface plot and residual plot for a long-term trend of volatility without the LAR method while Figure \ref{fig3} (c and d) shows the surface plot and residual plot for the long term trend with LAR method. It can be seen that, for training data sets, the long-term trend has better results with LAR (RMSE = $0.01615$) than without LAR (RMSE = $0.0745$) methods. The prediction results also highlight that the long-term trend output with LAR (RMSE = $0.10087$) is better than without LAR (RMSE = $0.203775$) methods.\\

Figure \ref{fig4} (a and b) and Figure \ref{fig4} (c and d) show the surface plots and residual plots for a seasonal component of volatility without LAR and with LAR methods. For training data sets, the LAR method (RMSE = $0.001123$) gives better results as compared to those without the LAR method (RMSE = $7.054$). The prediction results are also best achieved using the LAR method (RMSE = $0.010343$) as compared to the without LAR method (RMSE = $10.7434$).\\

Figure \ref{fig5} (a and b) shows the surface plot and residual plot for the remainder fluctuation of volatility without the LAR method. Figure \ref{fig5} (c and d) shows the surface plot and residual plots for the remainder fluctuation of volatility with the LAR method. For training data sets, the LAR method (RMSE = $0.0001294$) gives better results as compared to those without the LAR method (RMSE = $0.4691$). The prediction results also highlight that the LAR method (RMSE = $0.014783$) gives better results as compared to those without the LAR method (RMSE = $14.9627$). \\


Volatility prediction is an important task to study market fluctuations. In the present work, the robust least squares method has been used in two ways, firstly without the LAR method and secondly with the LAR method. This study has presented the general prediction equations for volatility and its three components. Most of the research to date has presented work on volatility only. For the first time, it is deduced that any variation in the three volatility components results in the variation of volatility itself. The authors also comment that the derived general equations can be directly used by young researchers to understand their data sets.\\
\section{Conclusions}
This paper focuses on proposing a simple, efficient, and more effective volatility prediction technique. Volatility prediction helps in a better understanding of market fluctuations over a period of time. But, the lack of available information due to irregularity makes the prediction study a challenging task. We have considered the daily CBOE volatility return price index for the prediction analysis. We have decomposed volatility into a long-term trend, seasonal, and remainder fluctuation components and then have applied the prediction method separately to each one of them. To compare it, we also predicted the volatility return indices. For the first time, the dependence of volatility on its three components viz. long-term trend, seasonal and remainder fluctuations have been discussed in detail. For prediction purposes, a simple regression analysis based on the robust least squares Curve Fitting method has been applied with two approaches viz. without LAR and with LAR methods. The obtained prediction results clearly depict that LAR based robust least squares method is acceptable for volatility prediction with the long-term trend, seasonal, and remainder fluctuation components. This study, for the first time, has provided general prediction equations using a simple approach for volatility and its components. The presented work can be extended in the future by applying artificial intelligence techniques to further improve the prediction results.
\section*{Acknowledgement}
The authors acknowledge the Yahoo Finance archive for providing the data \cite{bse}.

\bibliographystyle{elsarticle-num}
\bibliography{manus_131022}
\end{document}